\documentclass[12pt]{article}
\usepackage{epsfig}
\usepackage{cite}
\usepackage[polish]{babel}
\newcommand{\pc}[1]{ {\bf[}{\bf{#1}}{\bf]} }

\newcommand{\re}[1]{(\ref{#1})}
\newcommand{\ot}{\otimes}
\newcommand{\ddq}{ { {d \rule{15pt}{0pt}}\over {d\ln{Q^2}} } }
\newcommand{\msbar}{\overline{MS}}

\newcommand{\eq}{\begin{equation}}
\newcommand{\eqx}{\end{equation}}
\newcommand{\eqn}{\begin{eqnarray}}
\newcommand{\eqnx}{\end{eqnarray}}
\newcommand{\dt}{\Delta}

\newcommand{\nin}{\noindent}
\newcommand{\as}{\widetilde{\alpha_s}(Q^2)}

\newcommand{\dl}{$ln^2(1/x)\,$}
\newcommand{\gam}{$g_1^{\gamma}\,$}
\begin{document}
\nin
\begin{center}
{\Large
INSTYTUT FIZYKI J"ADROWEJ}\\

{\large
im. Henryka Niewodnicza"nskiego}\\

ul. Radzikowskiego 152, 31-342 Krak"ow \\

\vspace{12pt}

{\bf www.ifj.edu.pl/reports/2003.html}\\

Krak"ow, kwiecie"n 2003\\
\rule{15cm}{0.1cm}

\vspace{3cm}

{\bf Raport Nr} \\

\vspace{12pt}

{\large \bf Analiza spolaryzowanego
rozpraszania g"l"ebokonieelastycznego\\ 
w obszarze ma"lych warto"sci zmiennej x Bjorkena\\ 
z uwzgl"ednieniem resumacji poprawek
logarytmicznych $ln^2(1/x)$.}\\

\vspace{24pt}

B. Ziaja-Motyka \\

\vspace{12pt}

Rozprawa habilitacyjna-Habilitation thesis

\end{center}
%%%%%%%%%%%%%%%%%%%%%%%%%%%%%%%%%%%%%%%%%%%%%%%%%%%%%%%%%%%%%%%%%%%%%%%%%%%%%%
%
% ABSTRAKT
%
%%%%%%%%%%%%%%%%%%%%%%%%%%%%%%%%%%%%%%%%%%%%%%%%%%%%%%%%%%%%%%%%%%%%%%%%%%%%%%
\newpage
\nin
{\bf Abstract:}\\ \\
In this thesis we consider the polarized deep inelastic scattering in the
region of low values of Bjorken variable, $x$. For an accurate description
of this process including the logarithmic corrections, $\ln^2(1/x)$,
is required. These corrections resummed strongly influence
the behaviour of the spin structure functions and their moments.

We formulate the evolution equations for the unintegrated parton distributions
which include a complete resummation of the double logarithmic contributions,
$\ln^2(1/x)$.
Afterwards, these equations are completed with the standard LO and NLO
DGLAP evolution terms, in order to obtain the proper behaviour of the parton
distributions at moderate and large values of $x$.

The equations obtained are applied to the following observables and
processes: (i) to the nucleon structure function, $g_1$, in the polarized
deep inelastic scattering, (ii) to the structure function of the polarized
photon, $g_1^{\gamma}$, in the scattering of a lepton on a polarized photon,
and (iii) to the differential structure function, $x_J\,d^2g_1/dx_J\,dk_J^2$,
in the polarized deep inelastic scattering accompanied by a forward jet. Case
(iii) is proposed to be a test process for the presence and the magnitude
of the $\ln^2(1/x)$ contributions. For each process the consequences of
including the logarithmic corrections are studied in a detail.

After integrating out the structure function, $g_1$, the moments of
the nucleon structure function are obtained. The contribution of the region
of low $x$ to these moments is estimated, and then discussed in the context of
the spin sum rules.

Finally, some predictions for the observables, the asymmetry and
the cross sections, in the processes (i)-(iii) are given.
They are important to planned
experiments with the polarized HERA and linear colliders, which
will probe the region of low values of Bjorken $x$.
\newpage
\nin
{\bf Streszczenie:}\\ \\
W tej rozprawie rozwa"ramy proces spolaryzowanego rozpraszania
g"l"ebokonieelastycznego w obszarze ma"lych warto"sci zmiennej $x$ Bjorkena.
Dla uzyskania pe"lnego opisu tego procesu niezb"edne jest uwzgl"ednienie
wk"lad"ow podw"ojnie logarytmicznych $\sim \ln^2(1/x)$. Te zresumowane wk"lady
daj"a znacz"ace poprawki do spinowych funkcji struktury i ich moment"ow.

W pracach, wchodz"acych w sk"lad tej rozprawy, sformu"lowali"smy r"ownania
ewolucji dla nieprzeca"lkowanych rozk"lad"ow parton"ow. R"ownania te
zawiera"ly pe"lny, zresumowany wk"lad poprawek podw"ojnie logarytmicznych.
A"reby uzyska"c poprawne zachowanie rozk"lad"ow parton"ow przy wi"ekszych
warto"sciach zmiennej $x$, do r"owna"n tych do"l"aczyli"smy cz"lony
standardowej ewolucji DGLAP w przybli"reniu wiod"acym i podwiod"acym.

Otrzymane r"ownania zastosowali"smy do opisu nast"epuj"acych obserwabli i
proces"ow: (i) funkcji struktury nukleonu $g_1$ w spolaryzowanym rozpraszaniu
g"l"ebokonieelastycznym, (ii) funkcji struktury spolaryzowanego fotonu,
$g_1^{\gamma}$, w rozpraszaniu leptonu na spolaryzowanym fotonie, (iii)
r"o"rniczkowej funkcji struktury $x_J\,d^2g_1/dx_J\,dk_J^2$ w spolaryzowanym
rozpraszaniu g"l"ebokonieelastycznym z produkcj"a pojedynczego p"eku cz"astek w
prz"od. Ostatni proces mo"re s"lu"ry"c jako test na obecno"s"c i wielko"s"c
wk"lad"ow podw"ojnie logarytmicznych. Dla ka"rdego z tych proces"ow
analizowali"smy szczeg"o"lowo wp"lyw w"l"aczenia poprawek typu $\ln^2(1/x)$
na uzyskane wyniki.

Po numerycznym przeca"lkowaniu funkcji struktury $g_1$ otrzymali"smy momenty
funkcji struktury nukleonu.  Szczeg"o"lowo przeanalizowali"smy
wk"lad do tych moment"ow,\\
pochodz"acy z obszaru ma"lych warto"sci zmiennej
$x$, tak"re w kontek"scie zachowania spinowych regu"l sum.

Wreszcie, podali"smy przewidywania dla niekt"orych obserwabli, kt"ore mog"a by"c
w przysz"lo"sci zmierzone w procesach (i)-(iii). Przewidywania te s"a istotne
dla przysz"lych eksperyment"ow w akceleratorze spolaryzowana HERA i
akceleratorach liniowych.
%%%%%%%%%%%%%%%%%%%%%%%%%%%%%%%%%%%%%%%%%%%%%%%%%%%%%%%%%%%%%%%%%%%%%%%%%%%%%%%
\newpage
%%%%%%%%%%%%%%%%%%%%%%%%%%%%%%%%%%%%%%%%%%%%%%%%%%%%%%%%%%%%%%%%%%%%%%%%%%%%%%%

%%%%%%%%%%%%%%%%%%%%%%%%%%%%%%%%%%%%%%%%%%%%%%%%%%%%%%%%%%%%%%%%%%%%%%%%%%%%%%%
%
% SPIS TRESCI
%
%%%%%%%%%%%%%%%%%%%%%%%%%%%%%%%%%%%%%%%%%%%%%%%%%%%%%%%%%%%%%%%%%%%%%%%%%%%%%%%
\tableofcontents

\newpage
%%%%%%%%%%%%%%%%%%%%%%%%%%%%%%%%%%%%%%%%%%%%%%%%%%%%%%%%%%%%%%%%%%%%%%%%%%%%%%%

\section{Wprowadzenie}

%%%%%%%%%%%%%%%%%%%%%%%%%%%%%%%%%%%%%%%%%%%%%%%%%%%%%%%%%%%%%%%%%%%%%%%%%%%%%%%
Zagadnienie identyfikacji sk"ladnik"ow spinu nukleonu jest jednym z najciekawszych
problem"ow fizyki cz"astek ostatnich lat
\cite{reya,jet3,vetterli,ioffe}. W roku 1988 kolaboracja
do"swiadczalna EMC uzyska"la dane, wskazuj"ace na to, "re udzia"l
sk"ladowych kwarkowych w spinie nukleonu jest bardzo ma"ly \cite{pemc}. Wyniki te
przeczy"ly oszacowaniom teoretycznym uzyskanym z regu"l sum Ellisa-Jaffe'go i Bjorkena
\cite{vetterli,ioffe}. Regu"ly te wi"aza"ly kwarkowe sk"ladniki spolaryzowanego nukleonu
ze sta"lymi $g_A$ i $g_8$ otrzymanymi w rozpadach $\beta$ neutronu i hyperonu.
Przy zaniedbaniu udzia"lu kwark"ow dziwnych $s$, szacowa"ly one wk"lad
kwark"ow do ca"lkowitego spinu nukleonu na oko"lo trzy pi"ate.
Wyra"zna rozbie"rno"s"c pomi"edzy danymi
do"swiadczalnymi a przewidywaniem teoretycznym zapocz"atkowa"la szereg
bada"n, po"swi"econych ''zagadce'' spinu nukleonu \cite{jet3}.

Od tego czasu powsta"lo wiele hipotez, proponuj"acych wyja"snienie
znikomego udzia"lu kwark"ow w spolaryzowanym nukleonie.
Oszacowanie udzia"lu kwark"ow przy u"ryciu regu"ly sum
Ellisa-Jaffe'go w pierwotnej postaci, to jest z zaniedbaniem wk"ladu
kwark"ow dziwnych, zosta"lo podwa"rone przez nowe dane eksperymentalne
potwierdzaj"ace wyniki kolaboracji EMC (zebrane np. w \cite{vetterli,jet3}).

Dalej, szczeg"o"lowa analiza spinowej funkcji struktury nukleonu $g_1$ wykonana
w przybli"reniu podwiod"acym (NLO) chromodynamiki kwantowej (QCD) wskaza"la na trudno"s"c
jednoznacznego okre"slenia sk"ladowej kwarkowej spolaryzowanego nukleonu.
Anomalia aksjalna \cite{axial,leader2,jet3}, maj"aca "zr"od"lo w tzw. diagramach tr"ojk"atnych,
powoduje, "re pierwszy moment singletowej sk"ladowej spinu nukleonu
$\dt \Sigma$, daj"acy wk"lad do regu"l sum, r"o"rni si"e o cz"lon proporcjonalny
do $\alpha_S \dt g$ w r"o"rnych schematach renormalizacyjnych NLO QCD. Poniewa"r
sk"ladowa gluonowa $\dt g$ jest rz"edu $O(\alpha_S^{-1})$, cz"lon ten jako
ca"lo"s"c zachowuje si"e jak $O(\alpha_S^0)$ i nie znika nawet dla du"rych warto"sci
przekazu p"edu $Q^2$.

Niewyja"sniony do ko"nca jest r"ownie"r udzia"l sk"ladowej orbitalnej momentu
p"edu $L$ w
ca"lkowitym spinie nukleonu \cite{jet3}. Pojawia si"e tu trudno"s"c
z jednoznacznym zdefiniowaniem orbitalnego momentu p"edu. W spinowej regule sum:
$\displaystyle {1 \over 2}=J_q+J_g$ ca"lkowite momenty p"edu kwark"ow
$J_q$ i gluon"ow s"a jednoznacznie okre"slone i
niezmiennicze ze wzgl"edu na cechowanie. Jednak rozdzia"l
tych moment"ow na sk"ladniki orbitalne i spinowe:
\eqn
J_q&=&{1 \over 2} \dt \Sigma + L_q \nonumber\\
J_g&=&\dt g +L_g
\eqnx
nie jest jednoznaczny i zale"ry od wyboru cechowania.

Prace, kt"ore s"a podstaw"a niniejszej rozprawy badaj"a spolaryzowane
rozpraszanie g"l"ebokonieelastyczne w obszarze ma"lych warto"sci
zmiennej $x$ Bjorkena. Wi"a"re si"e to r"ownie"r z jedn"a z hipotez na temat
rozbie"rno"sci danych do"swiadczalnych i przewidywa"n teoretycznych dla spinu
nukleonu.
Do tej pory eksperymenty mierz"ace momenty spinowej funkcji
struktury nukleonu penetrowa"ly obszar ograniczony minimaln"a warto"sci"a
zmiennej $x$, $x\sim 10^{-3}$ \cite{smc98}.
Analiza teoretyczna sugeruje jednak, "re obszar
mniejszych $x$, $x<10^{-3}$ mo"re by"c kluczowy dla poprawnego wyznaczenia
warto"sci tych moment"ow, w tym zakresie $x$ zaczynaj"a bowiem dominowa"c
wk"lady pochodz"ace z resumacji podw"ojnych logarytm"ow $ln^2(1/x)$.
Mog"a one generowa"c du"re poprawki do funkcji struktury $g_1$
\cite{BARTNS,BARTS,blum9603,blum9606,kiyo,BBJK,kz}
i, dalej,
du"re wk"lady do moment"ow tej"re funkcji struktury $g_1$
\cite{BARTNS,BARTS,kz,blum95,blum9603,blum9606,blum99}.
Tak du"re wk"lady z obszaru asymptotycznego mog"lyby w konsekwencji 
zmieni"c dotychczasowe oszacowania eksperymentalne udzia"lu parton"ow w spinie
nukleonu.

W cyklu prac:

{\footnotesize
{\bf I} J. Kwieci"nski, B. Ziaja, ''QCD expectations for the spin structure
function $g_1$ in the low $x$ region'', Phys. Rev. D60:054004, 1999,\\

{\bf II} J. Kwieci"nski, B. Ziaja, ''Polarized deep inelastic scattering
accompanied by a forward jet as a probe of the $ln^2(1/x)$ resummation'',
Phys. Lett. B464:293, 1999,\\

{\bf III} J. Kwieci"nski, B. Ziaja, ''QCD predictions for polarized deep
inelastic scattering accompanied by a forward jet in the low $x$ region of
possible HERA measurements'', Phys. Lett. B470:247, 1999,\\

{\bf IV} J. Kwieci"nski, B. Ziaja, ''QCD predictions for spin dependent photonic
structure function $g_1^{\gamma}(x, Q^2)$ in the low $x$ region of future linear
colliders'', Phys. Rev. D63:054022, 2001,\\

{\bf V} B. Ziaja, ''Low $x$ double $ln^2(1/x)$ resummation effects at the sum
rules for nucleon structure function $g_1$'', Acta Phys. Polon. B32:2863, 2001,\\

{\bf VI} B. Bade"lek, J. Kwieci"nski, B. Ziaja, ''Spin structure function $g_1(x,
Q^2)$ and the DHGHY integral $I(Q^2)$ at low $Q^2$: predictions from the GVMD
model'', Eur.\ Phys.\ J.\ C{\bf 26}:45, 2002,\\

{\bf VII} B. Ziaja, ''Double logarithms, $ln^2(1/x)$, and the NLO DGLAP evolution
for the nonsinglet component of the nucleon spin structure function $g_1$'',
Phys. Rev. D66:114017, 2002,\\

{\bf VIII} B. Ziaja, ''Proton spin structure function, $g_1$, with the unified
evolution equations including NLO DGLAP terms and double logarithms,
$ln^2(1/x)$'', przyj"ete do druku w Eur. Phys. J C, 2003
}

\noindent
zosta"l zbadany wp"lyw resumacji poprawek podw"ojnie logarytmicznych,
$\sim ln^2(1/x)$ na zachowanie spinowej funkcji struktury nukleonu $g_1$
otrzymanej w rozpraszaniu\\
g"l"ebokonieelastycznym (DIS) oraz na regu"ly sum, kt"ore ta funkcja
spe"lnia (prace \pc{I,V-VIII}). W pracach tych zosta"ly sformu"lowane
zunifikowane r"ownania ewolucji,\\
uwzgl"edniaj"ace resumacj"e podw"ojnych logarytm"ow w obszarze ma"lych
warto"sci zmiennej $x$ Bjorkena i standardow"a ewolucj"e
Dokshitzer-Gribov-Lipatov-Altarelli-Parisi (DGLAP) dla "srednich i du"rych
warto"sci $x$.

R"ownania te pos"lu"ry"ly r"ownie"r do zbadania
fotonowej funkcji struktury $g_1^{\gamma}$ w obszarze ma"lych
warto"sci zmiennej $x$ (praca [{\bf IV}]). Wyniki uzyskane dla $g_1^{\gamma}$
s"a istotne dla projektowanych
zderzaczy liniowych $e^+e^-$ i $e\gamma$.

Zaproponowany zosta"l tak"re test
do"swiadczalny na obecno"s"c i wielko"s"c poprawek logarytmicznych w obszarze
ma"lych warto"sci $x$ (prace [{\bf II, III}]). W analogii do przypadku
niespolaryzowanego \cite{mueller1,mueller2},
pos"lu"ry"l do tego proces rozpraszania\\
g"l"ebokonieelastycznego z produkcj"a pojedynczego p"eku cz"astek w prz"od.
W uzyskanych przewidywaniach na wielko"sci mierzalne
zosta"ly uwzgl"ednione ci"ecia
kinematyczne\\ planowane dla przysz"lych pomiar"ow spolaryzowanego
rozpraszania\\ g"l"ebokonieelastycznego w akceleratorze HERA (praca \pc{III}).
%%%%%%%%%%%%%%%%%%%%%%%%%%%%%%%%%%%%%%%%%%%%%%%%%%%%%%%%%%%%%%%%%%%%%%%%%%%%%

\section{Struktura spolaryzowanego nukleonu w obszarze ma"lych warto"sci
zmiennej x Bjorkena}

%%%%%%%%%%%%%%%%%%%%%%%%%%%%%%%%%%%%%%%%%%%%%%%%%%%%%%%%%%%%%%%%%%%%%%%%%%%%%

%%%%%%%%%%%%%%%%%%%%%%%%%%%%%%%%%%%%%%%%%%%%%%%%%%%%%%%%%%%%%%%%%%%%%%%%%%%%%%%%
\subsection{Spinowa funkcja struktury nukleonu $g_1(x,Q^2)$}
%%%%%%%%%%%%%%%%%%%%%%%%%%%%%%%%%%%%%%%%%%%%%%%%%%%%%%%%%%%%%%%%%%%%%%%%%%%%%%%%

%%%%%%%%%%%%%%%%%%%%%%%%%%%%%%%%%%%%%%%%%%%%%%%%%%%%%%%%%%%%%%%%%%%%%%%%%%
\noindent
\begin{figure}[t]
\begin{center}
\epsfig{width=6cm, file=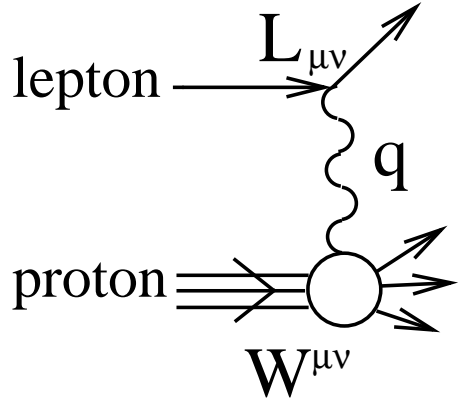}
\end{center}
\caption{Spolaryzowane rozpraszanie g"l"ebokonielastyczne leptonu na protonie
-najprostszy diagram}
\label{f1}
\end{figure}
%%%%%%%%%%%%%%%%%%%%%%%%%%%%%%%%%%%%%%%%%%%%%%%%%%%%%%%%%%%%%%%%%%%%%%%%%%%%%%%%
Rozwa"ramy spolaryzowane rozpraszanie g"l"ebokonieelastyczne leptonu
na nukleonie \cite{reya}.
Najprostszy proces tego typu opisany jest diagramem jak na Rys.\ \ref{f1}.
Zak"ladamy,
"re w stanie pocz"atkowym  zar"owno lepton jak i nukleon s"a spolaryzowane.
Wtedy przekr"oj czynny na to rozpraszanie wyra"rony jest poprzez zw"e"renie
iloczynu dw"och tensor"ow: tensora leptonowego $L_{ \mu\nu }$ i tensora
hadronowego $W^{\mu\nu}$, $L_{ \mu\nu }W^{\mu\nu}$ \cite{reya,ioffe}
(Rys.\ \ref{f1}). Tensor $W^{\mu\nu}$ mo"re by"c rozwini"ety na niezmienniki Lorentza.
Wsp"o"lczynniki tego rozwini"ecia definiuj"a mierzalne funkcje struktury.
Funkcja struktury $g_1$ jest wsp"o"lczynnikiem przy asymetrycznym cz"lonie
typu $\epsilon^{\mu\nu\rho\sigma} q_{\rho}S_{\sigma}$, gdzie $q$ jest
czterowektorem przekazu p"edu, a $S$ wektorem polaryzacji nukleonu w stanie
pocz"atkowym.

W eksperymencie do wyznaczenia warto"sci $g_1$ u"rywa si"e asymetrii
$A=(\sigma^{\uparrow\downarrow}-\sigma^{\uparrow\uparrow})/
(\sigma^{\uparrow\downarrow}+\sigma^{\uparrow\uparrow})$ \cite{reya,ioffe},
gdzie wsp"o"lczynniki
$\sigma$ oznaczaj"a ca"lkowite przekroje czynne w przypadku, gdy nukleon i lepton
s"a spolaryzowane zgodnie ($\sigma^{\uparrow\uparrow}$) lub przeciwnie
($\sigma^{\downarrow\uparrow}$).
Zwi"azek asymetrii z funkcj"a $g_1$ wymaga r"ownie"r znajomo"sci
niespolaryzowanej funkcji struktury $F_1$ \cite{reya,ioffe}.

W obszarze ma"lych warto"sci zmiennej $x$ Bjorkena asymptotyka funkcji $g_1$
przewidywana przez model biegun"ow Regge'go daje s"labe skalowanie $g_1$ z $x$
\cite{regg1,regg2}:
\eq
g_1^{i}(x,Q^2) = \gamma_i(Q^2)x^{-\alpha_{i}(0)},
\label{rg1}
\eqx
gdzie $g_1^{i}(x,Q^2)$ jest singletow"a ($i=S$) lub niesingletow"a ($i=NS$)
sk"ladow"a $g_1$. Oczekuje si"e, "re wsp"o"lczynnik Regge'go $\alpha_{i}(0)$ dla singletowej
i niesingletowej sk"ladowej $g_1$ jest podobny i bliski $0$:
$\alpha_{S,NS}(0)\leq 0$ \cite{kz}.

Posta"c asymptotyczna $g_1$ otrzymana ze standardowej ewolucji DGLAP
QCD daje w granicy
$x\rightarrow0$ zachowanie bardziej osobliwe ni"r w modelu Regge'go
(por.\ \cite{BARTNS,BARTS}).

Jednakowo"r, obydwa te przewidywania nie uwzgl"edniaj"a wk"ladu
logarytm"ow $ln(1/x)$, kt"ory staje si"e znacz"acy dla ma"lych warto"sci $x$.
Resumacja logarytm"ow $ln(1/x)$ w przypadku niespolaryzowanym zachodzi
poprzez r"ownanie BFKL \cite{bfkl1,bfkl2}. W przypadku spolaryzowanym analiza
teoretyczna \cite{BARTNS,BARTS} wykaza"la, "re dominuj"a wk"lady podw"ojnie
logarytmiczne \dl, generowane przez diagramy drabinkowe i pozadrabinkowe typu
''bremsstrahlung''. Diagramy drabinkowe odpowiadaj"a diagramom
 rozpraszania w prz"od fotonu na protonie z wygenerowan"a radiacyjnie
drabin"a partonow"a \cite{BARTNS,BARTS,BBJK}.
Diagramy te resumuj"a j"adra ewolucji DGLAP: $\dt P(z)$
w rz"edzie wiod"acym (LO) przy zerowych
warto"sciach u"lamku p"edu pod"lu"rnego $z=0$.
Diagramy pozadrabinkowe odpowiadaj"a diagramom rozpraszania w prz"od
z dodatkowymi propagatorami\\ partonowymi, do"l"aczonymi do r"o"rnych
segment"ow drabiny radiacyjnej \cite{QCD,QCD1}.

W obydwu przypadkach dla uzyskania wk"lad"ow podw"ojnie logarytmicznych
kluczowe jest uporz"adkowanie
p"ed"ow wymienianych parton"ow \cite{BARTNS,BARTS}.
Parametrem\\
uporz"adkowania jest stosunek
kwadratu p"edu poprzecznego partonu $k_n$ i u"lamku jego p"edu pod"lu"rnego
$x_n$, $k_n^2/x_n$. U"lamek p"edu $x_n$ odpowiada zmiennej
$\beta$ Sudakowa (por. \cite{BARTS}).
W przypadku diagram"ow pozadrabinkowych dodatkowym warunkiem jest,
aby do"l"aczone partony nios"ly p"ed mniejszy od
p"ed"ow wymienianych przez partony drabinkowe w danym segmencie drabiny.
Resumacja diagram"ow pozadrabinkowych\\ nast"epuje poprzez podczerwone
r"ownanie ewolucji \cite{QCD,QCD1}.

Szczeg"o"lowa analiza poprawek podw"ojnie logarytmicznych (DL)
zosta"la przeprowadzona w pracach \cite{BBJK} i \pc{I}. W obydwu pracach
resumacj"e tych poprawek wykonano dla nieprzeca"lkowanych rozk"lad"ow
parton"ow.
Zaproponowane zosta"lo ca"lkowe r"ownanie ewolucji sumuj"ace wk"lady drabinkowe i
pozadrabinkowe dla cz"e"sci singletowej i niesingletowej funkcji struktury
$g_1$. R"ownanie to przetransformowane do przestrzeni moment"ow poprzez
transformat"e Mellina poprawnie odtwarza"lo wymiary anomalne otrzymane
z podczerwonych r"owna"n ewolucji \cite{QCD,QCD1}.

Nale"ry podkre"sli"c, "re sformu"lowanie r"owna"n ewolucji w j"ezyku
nieprzeca"lkowanych rozk"lad"ow parton"ow umo"rliwi"lo "latw"a resumacj"e
poprawek logarytmicznych. Otrzymane r"ownania maj"a posta"c r"owna"n ca"lkowych z cz"lonem
niejednorodnym, kt"orym jest tzw. ''input'' nieperturbacyjny. Ewolucja \dl
w"l"aczona jest w j"adra r"ownania znajduj"ace si"e w cz"lonach jednorodnych.
Taka posta"c r"owna"n umo"rliwia r"ownie"r proste do"l"aczenie
cz"lon"ow standardowej ewolucji DGLAP, co jest niezb"edne dla
uzyskania prawid"lowego zachowania rozk"lad"ow parton"ow w obszarze "srednich
i du"rych warto"sci $x$ \pc{I}.

Pe"lna posta"c zunifikowanych r"owna"n dla niesingletowego nieprzeca"lkowanego\\
rozk"ladu kwark"ow $f_{NS}$ jest nast"epuj"aca (por.\ \pc{VII}):
\eqn
f_{NS}(x,Q^2)&=&\as(\dt P \ot \dt q_{NS}^{(0)})(x)
+\as\int_{k_0^2}^{Q^2} {dk^2 \over k^2}\,(\dt P_{reg}\ot f_{NS})(x,k^2)\nonumber\\
&&\hspace*{10ex}{\bf(\hspace*{3ex}DGLAP\hspace*{3ex})}\nonumber\\
&+&
\as{4\over 3}
\int_{x}^1 {dz\over z}
\int_{Q^2}^{Q^2/z}
{dk^{2}\over k^{2}}
f_{NS}\left({x\over z},k^{2}\right)\nonumber\\
&&\hspace*{10ex}{\bf (\hspace*{3ex}Ladder\hspace*{3ex})}\nonumber\\
&-&\as
\int_{x}^1 {dz\over z}
\Biggl(
\Biggl[ \frac{\tilde  {\bf F}_8 }{\omega^2} \Biggr](z)
\frac{ {\bf G}_0 }{2\pi^2}
\Biggr)_{qq}
\int_{k_0^2}^{Q^2}
{dk^{2}\over k^{2}}
f_{NS}\left({x\over z},k^{2}\right)\nonumber\\
&-&\as
\int_{x}^1 {dz\over z}
\int_{Q^2}^{Q^2/z}
{dk^{2}\over k^{2}}
\Biggl(
\Biggl[\frac{\tilde   {\bf F}_8 }{\omega^2} \Biggr]
\Biggl(\frac{k^{2}}{Q^2}z \Biggr)\frac{ {\bf G}_0 }{2\pi^2}
\Biggr)_{qq}
f_{NS}\left({x\over z},k^{2}\right).\nonumber\\
&&\hspace*{10ex}{\bf(Non-ladder)}
\label{nloinf}
\eqnx
Charakterystyka j"ader tego
r"ownania znajduje si"e w Dodatku A. Macierze ${\bf F}_8$ i ${\bf G}_0$
reprezentuj"a odpowiednio oktetow"a  fal"e cz"astkow"a i macierz czynnik"ow
kolorowych (Dodatek A).
Symbol $\displaystyle \Biggl[{\widetilde { {\bf F}_8}/{\omega^2} }\Biggr](z)$
oznacza odwrotn"a transformat"e Mellina wielko"sci
$\displaystyle {{\bf F}_8}/{\omega^2}$~:
\eq
\Biggl[{\widetilde { {\bf F}_8}/{\omega^2}  }\Biggr](z)=
\int_{\delta-i\infty}^{\delta+i\infty} {d\omega \over 2\pi i}
z^{-\omega}{{\bf F}_8(\omega)}/{\omega^2}.
\label{imellin}
\eqx
Kontur ca"lkowania przebiega na prawo od osobliwo"sci funkcji 
$\displaystyle {\bf F}_8(\omega)/{\omega^2}$.

Dla singletowego rozk"ladu $f_S$ r"ownanie \re{nloinf} zast"epowane jest
podobnym wektorowym r"ownaniem  ca"lkowym dla "l"acznej ewolucji
sk"ladowych singletowych : kwarkowej i gluonowej (por.\ \pc{VIII}).

Standardowe (przeca"lkowane) rozk"lady parton"ow $\dt q_i$, $\dt g$
($i=NS,S$) wi"a"r"a si"e z
nieprzeca"lkowanymi rozk"ladami w nast"epuj"acy spos"ob \cite{BBJK}, \pc{I}:
\eq
\dt q_i(x,Q^2)= \dt q_i^{(0)}(x)+\int_{k_0^2}^{W^2}\,{dk^2 \over k^2}\,
f_i(x^{\prime}=x(1+{k^2\over Q^2}),k^2).
\label{fix}
\eqx
W por"ownaniu do ewolucji DGLAP, przestrze"n fazowa zosta"la tu
rozszerzona z $Q^2$ do $W^2=Q^2(1/x-1)$ odpowiadaj"acego kwadratowi ca"lkowitej
energii mierzonej w uk"ladzie "srodka masy.

Ostatecznie, funkcj"e struktury $g_1$ otrzymuje si"e poprzez konwolucj"e przeca"lkowanych
rozk"lad"ow parton"ow ze wsp"o"lczynnikami Wilsona, kwarkowym $\dt C_q$ i
gluonowym $\dt C_g$ \cite{reya}:
\eqn
g_1(x,Q^2)&=&{1 \over 2}\,\sum_{q=1}^{N_f}\,e^2_q
\left\{\left(\dt C_q \ot (\dt q + \dt {\bar
q})\right)(x,Q^2)\right.\nonumber\\
&&\left.\rule{50pt}{0pt}+ (2 \dt C_g \ot \dt g)(x,Q^2)\right\},
\label{wilson}
\eqnx

W pracach \pc{I,VII,VIII} r"ownanie \re{nloinf} by"lo stopniowo udoskonalane.
Praca \pc{I}
uwzgl"ednia"la ewolucj"e DGLAP  tylko w rz"edzie wiod"acym (LO),
prace \pc{VII,VIII} rozszerzy"ly cz"e"s"c DGLAP o rz"ad podwiod"acy (NLO).
W"l"aczenie poprawek NLO by"lo nietrywialne z uwagi na konieczno"s"c
unikni"ecia podw"ojnego liczenia wk"lad"ow \dl\\ generowanych przez
cz"lony NLO DGLAP i przez wk"lady drabinkowe i pozadrabinkowe (Ladder,
Non-Ladder)
w przekrywaj"acych si"e obszarach przestrzeni fazowej\\ \pc{VII,VIII}~.
Aby to osi"agn"a"c, przestrze"n fazowa tych r"owna"n
zosta"la podzielona na dwa obszary: (i) $k_0^2<k^2<Q^2$ i
(ii) $Q^2<k^2<Q^2/z$. W obszarze (i) zachowano wszystkie cz"lony (osobliwe)
generowane przez cz"lony drabinkowe i pozadrabinkowe
i dodano tylko nieosobliwe cz"lony typu DGLAP: $\dt
P_{reg}(z)$ w rz"edzie wiod"acym i podwiod"acym, to znaczy cz"lony DGLAP
zbie"rne w granicy $z\rightarrow 0$. W obszarze (ii) cz"lony typu DGLAP
w og"ole nie wyst"epowa"ly i wk"lady logarytmiczne pochodzi"ly tylko od
cz"lon"ow drabinkowych i pozadrabinkowych.

Taka procedura w"l"aczenia cz"lon"ow NLO DGLAP do zunifikowanych r"owna"n ewolucji
jest jednoznaczna. Opiera si"e ona na twierdzeniu dowiedzionym w \cite{QCD,QCD1},
i"r resumacja podw"ojnych logarytm"ow jest kompletna po uwzgl"ednieniu
wszystkich wk"lad"ow od diagram"ow
drabinkowych i pozadrabinkowych. Jak pokazano w pracy \pc{VIII} poprawne
wyliczenie funkcji struktury wymaga u"rycia pe"lnego kwarkowego i gluonowego\\
wsp"o"lczynnika Wilsona z uwzgl"ednieniem jego cz"e"sci osobliwej
i nieosobliwej.

Przeprowadzona wed"lug powy"rszego schematu analiza sk"ladowej niesingletowej\\
funkcji $g_1$: $g_1^{NS}$
da"la nast"epuj"ace rezultaty. Wyniki uzyskane ze zunifikowanej ewolucji
nieperturbacyjnej parametryzacji wej"sciowej, 
uwzgl"edniaj"acej wk"lady podw"ojnie logarytmiczne
i standardow"a ewolucj"e DGLAP w rz"edzie wiod"acym (DL+LO) lub podwiod"acym
(DL+NLO) oraz wyniki uzyskane ze standardowej ewolucji NLO DGLAP r"o"rni"ly si"e
niewiele mi"edzy sob"a dla $x>10^{-3}$. R"o"rnice mi"edzy krzywymi z
r"o"rnych przybli"re"n by"ly rz"edu kilkunastu procent. Znacz"ace
rozbie"rno"sci pojawia"ly si"e w obszarze bardzo
ma"lych $x\leq 10^{-4}$, gdy dominowa"ly poprawki \dl. Wielko"s"c
tych rozbie"rno"sci zale"ra"la od u"rytej parametryzacji wej"sciowej
(inputu).

Podobna analiza przeprowadzona dla cz"e"sci singletowej $g_1$ i, w konsekwencji,
dla pe"lnej funkcji $g_1$ oraz dla rozk"ladu gluon"ow da"la zbli"rone
rezultaty na poziomie ewolucji LO i DL+LO \pc{I} i ewolucji NLO i DL+NLO
\pc{VIII}. Zw"laszcza dla rozk"ladu gluon"ow rozbie"rno"sci pomi"edzy
ewolucj"a DL+NLO(LO) a standardow"a ewolucj"a NLO(LO) by"ly du"re. Potwierdza
to, "re zunifikowana ewolucja rozk"ladu gluon"ow jest zdominowana przez wk"lad
diagram"ow drabinkowych i pozadrabinkowych.

W"l"aczaj"ac cz"lony podwiod"ace NLO w cz"e"sci DGLAP, zbadano
zale"rno"s"c uzyskanych wynik"ow od u"rytego schematu renormalizacji NLO.
Wybrano dwa schematy renormalizacji: $\msbar$ \cite{vann1,vogel}
i JET/CI \cite{jet4,axial,jet2,jet3,cheng}
nale"r"ace do jednej rodziny  schemat"ow typu $\msbar$
\cite{vann2,leader1,leader2}. W obr"ebie tej grupy
niesingletowy rozk"lad kwark"ow $\dt q_{NS}$ i rozk"lad gluon"ow $\dt g$ s"a
niezale"rne od wyboru schematu, kt"ory zmienia tylko singletowy
rozk"lad kwark"ow $\dt \Sigma$ \cite{vann2,leader1,leader2}.
Zmiany schematu renormalizacji dokonali"smy poprzez u"rycie w zunifikowanych
r"ownaniach j"ader ewolucji DGLAP w"la"sciwych dla danego schematu oraz
odpowiedni"a transformacj"e parametryzacji wej"sciowej rozk"ladu kwark"ow.

Badaj"ac transformacj"e do schematu JET/CI zauwa"ryli"smy, "re rozwini"ecie
cz"lon"ow transformuj"acych w sta"lej sprz"e"renia $\alpha_S$ i obci"ecie tych
cz"lon"ow z dok"ladno"sci"a do ustalonego rz"edu w $\alpha_S$ nie jest w"la"sciwe
w obszarze ma"lych warto"sci $x$. W tym obszarze logarytmy $ln(1/x)$
osi"agaj"a du"re warto"sci i w"la"sciwe rozwini"ecie perturbacyjne powinno
uwzgl"ednia"c cz"lony rz"edu $\alpha_S^n\,\ln^m(1/x)$, a nie tylko cz"lony 
rz"edu $\alpha_S^n$.

Aby unikn"a"c tych problem"ow, w r"ownaniach transformuj"acych ze schematu
$\msbar$ do schematu JET/CI ograniczyli"smy si"e do j"ader ewolucji $\msbar$
DGLAP z dok"ladno"sci"a do cz"lon"ow wiod"acych (LO). Dzi"eki temu wyliczyli"smy
analitycznie pe"ln"a posta"c transformacji mi"edzy schematami, uwzgl"edniaj"ac"a
wszystkie rz"edy rozwini"ecia w $\alpha_S$. W tym przypadku by"l to rz"ad pod-podwiod"acy
(NNLO). Otrzymana ewolucja NNLO w schemacie JET/CI odpowiada"la ewolucji
LO DGLAP w schemacie $\msbar$.

W"l"aczyli"smy ewolucj"e NNLO DGLAP w schemacie JET/CI do zunifikowanych
r"owna"n i otrzymali"smy wyniki dla $g_1$ oraz $\dt g$.
Por"ownuj"ac je z wynikami zunifikowanej ewolucji w schemacie $\msbar$,
zaobserwowali"smy znacz"acy wzrost bezwzgl"ednej warto"sci funkcji $g_1$
otrzymanej z ewolucji DL+NNLO (JET) w stosunku do $g_1$ otrzymanej z ewolucji
NNLO (JET). Efekt ten wywo"lany by"l dodaniem w sektorze singletowym
du"rych, ujemnych wk"lad"ow z ewolucji drabinkowej i pozadrabinkowej
oraz du"rych, ujemnych wk"lad"ow z r"ownania transformacji mi"edzy schematami.

Wyniki, uzyskane dla rozk"ladu gluon"ow by"ly natomiast podobne do wynik"ow\\
uzyskanych ze schematu $\msbar$. Potwierdza to obserwacj"e, i"r zunifikowana ewolucja
rozk"ladu gluon"ow jest zdominowana przez wk"lady drabinkowe i pozadrabinkowe.

Powy"rsze obserwacje wyra"znie wskazuj"a, "re standardowa ewolucja DGLAP jest
niepe"lna w obszarze ma"lych warto"sci $x$. Nie uwzgl"ednia ona wszystkich
wk"lad"ow z resumacji $\ln(1/x)$, kt"ore s"a w tym obszarze du"re i
nie powinny by"c pomijane.
%%%%%%%%%%%%%%%%%%%%%%%%%%%%%%%%%%%%%%%%%%%%%%%%%%%%%%%%%%%%%%%%%%%%%%%%%%%%%%%%
\subsection{Regu"ly sum dla spolaryzowanego nukleonu}
%%%%%%%%%%%%%%%%%%%%%%%%%%%%%%%%%%%%%%%%%%%%%%%%%%%%%%%%%%%%%%%%%%%%%%%%%%%%%%%%

Regu"ly sum s"a to relacje spe"lniane przez momenty funkcji struktury. Dla
rozpraszania g"l"ebokonieelastycznego otrzymuje si"e je z czasoprzestrzennej
reprezentacji urojonej cz"e"sci amplitudy rozpraszania $Im\, T^a_{\mu\nu}$
poprzez komutatory pr"adowe \cite{ioffe}:
\eq
Im\, T_{ik}(x)_{x_0\rightarrow 0}=-\epsilon_{ikl}\,
\langle p,s\mid \frac{1}{3}\left[j_{5l}^3+\frac{1}{\sqrt{3}}j_{5l}^8\right]
+\frac{2}{9}j_{5l}^0\mid p,s\rangle,
\label{rsum}
\eqx
gdzie $j_{5l}^3$, $j_{5l}^8$, $j_{5l}^0$ s"a odpowiednio aksjalnymi
pr"adami: izowektorowym, oktetowym i singletowym przy za"lo"reniu zapachowej
symetrii SU(3).

Warunek symetrii izospinowej, na"lo"rony na element macierzowy z prawej strony
r"ownania \re{rsum}, zawieraj"acy pr"ad izowektorowy,
wyra"ra go poprzez sta"l"a rozpadu $\beta$ neutronu $g_A$
\cite{ioffe,vetterli,reya}.

Warunek symetrii zapachowej SU(3) na"lo"rony na element macierzowy zawieraj"acy pr"ad oktetowy daje zwi"azek tego elementu
ze sta"lymi rozpadu $\beta$ barionu $F$ i $D$ \cite{ioffe,vetterli}.

Lew"a stron"e r"ownania \re{rsum} mo"rna wyrazi"c poprzez
kombinacj"e funkcji struktury nukleonu. Pozwala to zapisa"c powy"rsze
warunki w nast"epuj"acy spos"ob:
\eqn
\int_0^1 dx\,(g_1^{proton}(x,Q^2)-g_1^{neutron}(x,Q^2))&=&g_A/6\label{bjork}\\
\int_0^1 dx\,g_1^8=\int_0^1 dx\,(\dt u+\dt d -2 \dt s) &=&3F-D.\label{ej}\\
\eqnx
Warunki te s"a znane jako odpowiednio regu"la sum Bjorkena i regu"la sum
Ellisa-Jaffe'go.

Istnieje jeszcze trzecia regu"la sum, kt"ora
wi"a"re pierwszy moment funkcji struktury $g_1$ ze statycznym
w"lasno"sciami nukleonu: jego anomalnymi momentami magnetycznymi
w granicy fotoprodukcji $Q^2=0$. Relacj"e
t"e otrzymuje si"e ze zwi"azku dyspersyjnego amplitudy rozpraszania
fotonu na nukleonie ze spinow"a funkcj"a struktury nukleonu \cite{ioffe,ioffe2}.
Bior"ac pod uwag"e, "re to rozpraszanie poni"rej pewnej granicznej
warto"sci masy niezmienniczej $W<W_{prog}$ jest zdominowane przez rezonanse
barionowe \cite{ioffe,ioffe2},
powy"rsz"a regu"l"e sum dla momentu $I(Q^2)$ zdefiniowanego jako:
\eq
I(Q^2)=I_{res}(Q^2)+M\int_{\nu_{prog}(Q^2)}^{\infty}\,
\frac{d\nu}{\nu^2}\,g_1\left(x(\nu),Q^2\right)
\label{imom}
\eqx
mo"rna zapisa"c w nast"epuj"acej postaci:
\eq
I(0)=I_{res}(0)+M\int_{\nu_{prog}(0)}^{\infty}\,
\frac{d\nu}{\nu^2}\,g_1\left(x(\nu),0\right)=-\kappa_{nucl}^2/4.
\label{imom0}
\eqx
Regu"la ta znana jest jako regu"la sum Drell'a-Hearn'a-Gerasimova-Hosody-Yamamoto (DHGHY) \cite{gdh1,gdh2,gdh3}.
Symbol $\displaystyle \kappa_{nucl}$ oznacza moment anomalny nukleonu,
za"s $I_{res}(Q^2)$ opisuje wk"lad do $I(Q^2)$ pochodz"acy od rezonans"ow
barionowych.

W pracach \pc{V} analizowali"smy momenty funkcji struktury $g_1$
protonu i neutronu, otrzymane ze zunifikowanych r"owna"n ewolucji DL+LO poprzez
przeca"lkowanie funkcji struktury $g_1$ w obszarze $10^{-5}<x<1$ przy
warto"sciach przekazu p"edu $Q^2$, $2< Q^2 <15$ GeV$^2$.
Sprawdzili"smy analitycznie, "re (niesingletowe) r"ownania zachowuj"a
regu"ly sum Bjorkena i Ellisa-Jaffe'go, je"sli regu"ly te s"a spe"lniane
przez nieperturbacyjne wej"sciowe rozk"lady parton"ow
$\dt q^{(0)}$ (tzw. "input"). Natomiast, w r"ownaniach ewolucji dla cz"e"sci
singletowej momenty singletowego rozk"ladu kwark"ow $\dt \Sigma$ i
rozk"ladu gluon"ow zale"r"a od $Q^2$.

Otrzymane wyniki numeryczne wykazywa"ly dobr"a zgodno"s"c z danymi
eksperymentalnymi dla protonu, natomiast dla neutronu wyniki te le"ra"ly
poni"rej tych danych. Przyczyn"a by"la prawdopodobnie parametryzacja
wej"sciowa rozk"ladu parton"ow dla neutronu, kt"ora nie odtwarza"la dobrze danych
przy $Q_0^2=1$ GeV$^2$.

Udzia"l obszaru asymptotycznego $10^{-5}<x<10^{-3}$ w ca"lkowitym momencie
protonu wynosi"l oko"lo 2\%. Udzia"l obszaru asymptotycznego
w ca"lkowitym momencie neutronu oszacowali"smy na oko"lo 8\%. Wk"lad
ten wzrasta"l z rosn"acym $Q^2$, osi"agaj"ac dla neutronu
maksymalnie 10\%.

W pracy \pc{VII} sprawdzili"smy zachowanie regu"ly sum Bjorkena w zunifikowanych
r"ownaniach ewolucji z do"l"aczonymi cz"lonami DGLAP w rz"edzie podwiod"acym
NLO. Pierwszy moment $\Gamma_1^{Bjorken}=\Gamma_1^{proton}-\Gamma_1^{neutron}$
otrzymany by"l z numerycznego\\ przeca"lkowania rozwi"aza"n r"owna"n
ewolucji na $g_1$ w obszarze $10^{-4}<x<1$. Uwzgl"ednienie wk"ladu
asymptotycznego z obszaru $x<10^{-4}$ zmieni"loby ten wynik o oko"lo $1$ \%.
Moment ten nie odtwarza"l regu"ly sum
dok"ladnie, jednak por"ownany z przewidywaniami QCD dla regu"ly Bjorkena
w rz"edach NLO i NNNLO wykazywa"l mniejsz"a rozbie"rno"s"c z
przewidywaniami NNNLO, odtwarzaj"ac dobrze kszta"lt krzywej NNNLO.
To sugeruje, "re zunifikowana ewolucja DL+NLO wykracza poza standardow"a
analiz"e NLO DGLAP: poprawki podw"ojnie logarytmiczne wygenerowane przez
cz"lony DL oddzia"lywuj"a silnie na zachowanie funkcji struktury w
obszarze ma"lych warto"sci $x$, co wp"lywa tak"re na zachowanie jej
moment"ow.

W pracy \pc{VI} wykorzystali"smy regu"l"e sum DHGHY \cite{gdh1,gdh2,gdh3}
dla otrzymania\\
przewidywa"n teoretycznych dla funkcji struktury $g_1$ w
obszarze ma"lych warto"sci przekazu
p"edu $Q^2$. Do parametryzacji funkcji struktury u"ryli"smy
uog"olnionego modelu dominacji mezon"ow wektorowych \cite{gvmd}.
Regu"la sum DHGHY pos"lu"ry"la tu dla ustalenia wielko"sci wk"ladu lekkich
mezon"ow do $g_1$. Otrzymane przewidywania dla funkcji $g_1$ oraz
dla momentu DHGHY $I(Q^2)$ (por. \re{imom}) zosta"ly por"ownane z danymi
eksperymentalnymi. Wykaza"ly z nimi zadowalaj"ac"a zgodno"s"c.
%%%%%%%%%%%%%%%%%%%%%%%%%%%%%%%%%%%%%%%%%%%%%%%%%%%%%%%%%%%%%%%%%%%%%%%%%%%%%%%%

\section{Struktura spolaryzowanego fotonu w zderzeniach $e^+e^-$, $e\gamma$}

%%%%%%%%%%%%%%%%%%%%%%%%%%%%%%%%%%%%%%%%%%%%%%%%%%%%%%%%%%%%%%%%%%%%%%%%%%%%%%%%
Zunifikowane r"ownania ewolucji, "l"acz"ace poprawki drabinkowe i
pozadrabinkowe oraz ewolucj"e DGLAP mo"rna tak"re stosowa"c do badania
struktury spolaryzowanego fotonu w rozpraszaniu g"l"ebokonieelastycznym
elektronu na fotonie \pc{IV}.

Spinowa funkcja struktury spolaryzowanego fotonu b"edzie dost"epna 
dla pomiar"ow eksperymentalnych w przysz"lych akceleratorach liniowych 
$e^+e^-$ i $e\gamma$ \cite{STRATMANN1,phot1,phot2}.\\
Rozpraszanie $e\gamma$ czyli rozpraszanie g"l"ebokonieelastyczne elektronu
lub pozytronu na strumieniu foton"ow jest szczeg"olnie odpowiednie do badania
struktury fotonu w obszarze ma"lych warto"sci zmiennej $x$ Bjorkena. Zawarto"s"c
parton"ow w spolaryzowanym fotonie b"edzie mo"rna r"ownie"r bada"c w procesach
dwup"ekowej fotoprodukcji\\ przy rozpraszaniu $ep$ w akceleratorze HERA
\cite{phot3}.

Adaptacja zunifikowanych r"owna"n do opisu rozpraszania elektronu na fotonie\\
wymaga w"l"aczenia do tych r"owna"n dodatkowych cz"lon"ow niejednorodnych
$\dt k_i(x,Q^2)$, opisuj"acych punktowe sprz"e"renie fotonu do kwark"ow,
antykwark"ow i gluon"ow \cite{STRATMANN1}.
Cz"lony te do"l"aczaj"a do ewolucji DGLAP
w nast"epuj"acy spos"ob:
\eqn
\ddq \dt q_{i}(x,Q^2)&=&\dt k_i(x,Q^2) + \as\,(\dt P \ot \dt q_{i})(x,Q^2),
\label{diff}
\eqnx
gdzie $i=S,NS,g$. W rz"edzie wiod"acym LO ewolucji DGLAP sprz"e"renie
punktowe foton-gluon nie wyst"epuje, $\dt k_g(x,Q^2)=0$ \cite{STRATMANN1}.

W pracy \pc{IV} znale"zli"smy najpierw rozwi"azanie analityczne uproszczonych
r"owna"n, opisuj"acych ewolucj"e DL w przybli"reniu drabinkowym. Dzi"eki
temu otrzymali"smy posta"c asymptotyczn"a funkcji \gam w granicy
$x\rightarrow0$. Zgodnie z oczekiwaniami, zale"rno"s"c ta jest pot"egowa z
ujemnym wyk"ladnikiem, pozostaj"acym w prostym zwi"azku z asymptotycznymi
wyk"ladnikami dla funkcji struktury protonu $g_1$.

Nast"epnie rozwi"azali"smy numerycznie pe"lne r"ownanie zunifikowane dla fotonu
z uwzgl"ednieniem ewolucji DGLAP w rz"edzie wiod"acym. Na wej"sciowe rozk"lady
parton"ow na"lo"ryli"smy warunek wynikaj"acy z regu"ly sum dla pierwszego
momentu funkcji \gam, \cite{bass1,bass2}:
\eq
\int_0^1\, dx\, g_1^{\gamma}(x,Q^2)=0.
\label{gamma0}
\eqx
Rozwa"ryli"smy dwa przypadki graniczne dla parametryzacji wej"sciowych
rozk"lad"ow parton"ow: (i) przypadek, kiedy zar"owno $\dt q^{(0)}$ jak i $\dt g^{(0)}$
s"a r"owne $0$ i rozwi"azania r"ownania ewolucji s"a generowane radiacyjnie
z cz"lon"ow niejednorodnych $\dt k$, (ii) przypadek, kiedy $\dt
q^{(0)}=0$, ale
rozk"lad gluon"ow jest niezerowy. Rozk"lad ten otrzymuje si"e
z modelu dominacji mezon"ow wektorowych przy za"lo"reniu dominacji mezon"ow
$\rho$ i $\omega$ \cite{vmd1,vmd2}.

Uwzgl"ednienie poprawek podw"ojnie logarytmicznych mia"lo widoczny wp"lyw na otrzymane
przewidywania dla \gam. Wielko"s"c r"o"rnic pomi"edzy krzywymi otrzymanymi
z analizy LO DGLAP i NLO DGLAP oraz DL+LO silnie  zale"ry od u"rytej
parametryzacji
wej"sciowej, a tak"re od wielko"sci obci"ecia p"edowego w podczerwieni $k_0^2$.
Jednak"re otrzymana  warto"s"c asymetrii $g_1^{\gamma}/F_1^{\gamma}$ jest bardzo
ma"la, rz"edu $10^{-3}$ dla $x=10^{-4}$, a wi"ec trudna do zmierzenia.

Uzyskany rozk"lad gluon"ow $\dt g$ zale"ry silnie od u"rytego
inputu. W przypadku (i) rozk"lad $\dt g$ staje si"e ujemny przy ma"lych
warto"sciach zmiennej $x$, natomiast w przypadku (ii) jest on dodatni.
W tej sytuacji
przysz"le pomiary b"ed"a mog"ly "latwo wskaza"c, kt"ora z parametryzacji
wej"sciowych jest w"la"sciwa.
%%%%%%%%%%%%%%%%%%%%%%%%%%%%%%%%%%%%%%%%%%%%%%%%%%%%%%%%%%%%%%%%%%%%%%%%%%%%%%%%

\section{Produkcja pojedynczego p"eku cz"astek\\ w rozpraszaniu
g"l"ebokonieelastycznym jako test\\ wk"ladu
poprawek logarytmicznych $ln^2(1/x)$}

%%%%%%%%%%%%%%%%%%%%%%%%%%%%%%%%%%%%%%%%%%%%%%%%%%%%%%%%%%%%%%%%%%%%%%%%%%%%%%%%
Procesem, kt"ory mo"re testowa"c wielko"s"c poprawek podw"ojnie logarytmicznych
\dl\\ jest proces spolaryzowanego rozpraszania g"l"ebokonieleastycznego
z produkcj"a pojedynczego p"eku cz"astek w prz"od.
Idea wykorzystania tego procesu jako testu na resumacj"e BFKL cz"lon"ow
logarytmicznych $ln(1/x)$ zosta"la pierwotnie sformu"lowana dla przypadku niespolaryzowanego
rozpraszania g"l"ebokonieelastycznego \cite{mueller1,mueller2}.

Za"l"o"rmy, "re w rozpraszaniu g"l"ebokonieelastycznym elektronu na protonie
produkowany jest pojedynczy p"ek cz"astek o p"edzie $(x_J,k_J^2)$, gdzie
$x_J$ jest u"lamkiem p"edu pod"lu"rnego niesionego przez p"ek, a $k_J^2$
oznacza kwadrat jego p"edu poprzecznego. Przekaz p"edu niesiony przez foton
wynosi $q$, a zmienna $x$ Bjorkena r"owna jest $x=Q^2/(2pq)$, gdzie
$Q^2=-q^2$ i $p$ jest p"edem protonu.

Je"reli wsp"o"lrz"edne p"edowe p"eku spe"lniaj"a za"lo"renia:
\eqn
x_J&\gg&x,\nonumber\\
k_J^2&\sim&Q^2,
\label{jet}
\eqnx
to p"ek produkowany jest w obszarze ma"lych warto"sci mierzalnej
zmiennej $x/x_J$, za"s potencjalny wk"lad do tej produkcji pochodz"acy od
standardowej ewolucji DGLAP jest t"lumiony ($k_J^2\sim Q^2$)
\cite{mueller1,mueller2,supp1,supp2,supp3,supp4}.
Warunek $k_J^2\sim Q^2$
zapewnia r"ownie"r ograniczon"a penetracj"e obszaru nieperturbacyjnego przez
ewolucj"e $ln(1/x)$ \cite{supp5,supp6}.

Wielko"sci"a mierzon"a w takim procesie jest r"o"rniczkowy przekr"oj czynny
wyra"rony poprzez przekaz p"edu $Q^2$, u"lamek energii elektronu niesiony
przez foton oddzia"lywania $y$ i wreszcie r"o"rniczkow"a funkcj"e
struktury $x_J\,d^2g_1/dx_J\,dk_J^2$.
W analogii do pe"lnej funkcji struktury $g_1$, r"o"rniczkow"a
funkcj"e struktury mo"rna wyrazi"c poprzez przeca"lkowane rozk"lady
parton"ow w protonie i nieprzeca"lkowane rozk"lady kwark"ow i antykwark"ow w
partonie \pc{II}. Dla nieprzeca"lkowanych rozk"lad"ow formu"luje si"e r"ownania
ewolucji podobne jak w przypadku spolaryzowanego rozpraszania
g"l"ebokonieelastycznego.\\ Ograniczaj"a si"e one tylko do wk"lad"ow drabinkowych
i pozadrabinkowych z uwagi
na t"lumienie ewolucji DGLAP (warunek \re{jet}) \pc{II}.

W pracy \pc{II} sformu"lowali"smy r"ownania ewolucji \dl dla nieprzeca"lkowanych
rozk"lad"ow kwark"ow i antykwark"ow w partonie. 
W r"ownaniach tych uwzgl"ednili"smy
wk"lady drabinkowe i pomin"eli"smy podwiod"ace wk"lady pozadrabinkowe.
R"ownania te rozwi"azali"smy analitycznie
dla dw"och przypadk"ow: (i) ustalonej sta"lej sprz"e"renia,
$\alpha_S(\mu^2)$, gdzie $\mu^2=(k_J^2+Q^2)/2$ \cite{supp1,supp2},
(ii) biegn"acej sta"lej
sprz"e"renia  $\alpha_S(\mu^2)$, gdzie $\mu^2=k_f^2/\zeta$ i $k_f^2$ oznacza
kwadrat sk"ladowej poprzecznej p"edu kwarku (antykwarku), za"s $\zeta$ jest
u"lamkiem jego p"edu pod"lu"rnego.
Po numerycznym przeca"lkowaniu\\ nieprzeca"lkowanych rozk"lad"ow dostali"smy
oszacowanie r"o"rniczkowej funkcji struktury $x_J\,d^2g_1/dx_J\,dk_J^2$.
Zale"ra"lo ono silnie od warto"sci zmiennej $x/x_J$, a tak"re od warto"sci
p"edu niesionego przez powsta"ly p"ek cz"astek $k_J^2$.
Silna zale"rno"s"c od warto"sci zmiennej $x/x_J$ jest bezpo"srednim
efektem uwzgl"ednienia resumacji podw"ojnych logarytm"ow.
Por"ownanie wynik"ow otrzymanych z u"ryciem ustalonej
i biegn"acej sta"lej sprz"e"renia wskaza"lo na ten ostatni przypadek
jako bardziej realistyczny. W obydwu przypadkach efekty resumacji DL
zdecydowanie wybija"ly si"e ponad t"lo, opisane przybli"reniem Borna.

W pracy \pc{III} oszacowali"smy przekr"oj czynny i
asymetri"e dla procesu produkcji p"eku cz"astek w prz"od w spolaryzowanym
rozpraszaniu g"l"ebokonieelastycznym. U"ryli"smy ci"e"c kinematycznych
przewidywanych dla akceleratora spolaryzowana HERA
\cite{dane1,dane2,dane3}.
Efekty resumacji \dl znacz"aco zwi"eksza"ly wielko"s"c asymetrycznego
przekroju czynnego $d\sigma/d x$, gdzie $x$ oznacza"lo zmienn"a Bjorkena.
Uzyskana warto"s"c asymetrii by"la jednak niewielka i zmienia"la si"e
mi"edzy $-0.01$ a $-0.04$ dla ma"lych warto"sci $x$.
%%%%%%%%%%%%%%%%%%%%%%%%%%%%%%%%%%%%%%%%%%%%%%%%%%%%%%%%%%%%%%%%%%%%%%%%%%%%%%%%

\section{Podsumowanie}

%%%%%%%%%%%%%%%%%%%%%%%%%%%%%%%%%%%%%%%%%%%%%%%%%%%%%%%%%%%%%%%%%%%%%%%%%%%%%%%%
Podsumowuj"ac, zestawiam najwa"rniejsze wyniki, uzyskane w pracach
\pc{I-VIII}.

 - Sformu"lowanie r"ownania ewolucji dla niesingletowej i 
singletowej sk"ladowej funkcji struktury
$g_1$ z uwzgl"ednieniem nietrywialnych poprawek pozadrabinkowych,
sumowanych podczerwonym r"ownaniem ewolucji \pc{I}. Sformu"lowanie zunifikowanego
r"ownania ewolucji z uwzgl"ednieniem cz"lon"ow wiod"acych i podwiod"acych
ewolucji DGLAP \pc{I,VII,VIII}. Otrzymanie przewidywa"n dla funkcji $g_1$ i
rozk"ladu gluon"ow
w obszarze ma"lych warto"sci zmiennej $x$ Bjorkena.

 - Otrzymanie przewidywa"n dla moment"ow spinowej funkcji struktury $g_1$\\
z uwzgl"ednieniem wk"ladu od resumacji podw"ojnych logarytm"ow \pc{V,VII}.
Otrzymanie
przewidywa"n dla funkcji struktury $g_1$ w obszarze ma"lych warto"sci $x$ i
przekazu p"edu $Q^2$ z wykorzystaniem regu"ly sum DHGHY \pc{VI}.

 - Zaproponowanie testu do"swiadczalnego na obecno"s"c i wielko"s"c poprawek
podw"ojnie logarytmicznych. Testem tym mo"re by"c proces produkcji pojedynczego
p"eku cz"astek w prz"od w spolaryzowanym rozpraszaniu g"l"ebokonieelastycznym
\pc{II,III}.

 - Zbadanie wp"lywu resumacji podw"ojnych logarytm"ow na zachowanie funkcji
struktury spolaryzowanego fotonu w obszarze ma"lych warto"sci $x$. Oszacowanie
wielko"sci mierzalnej: asymetrii. Oszacowanie udzia"lu gluon"ow w
fotonie \pc{IV}.

%%%%%%%%%%%%%%%%%%%%%%%%%%%%%%%%%%%%%%%%%%%%%%%%%%%%%%%%%%%%%%%%%%%%%%%%%%%%%%%%

\section*{Podzi"ekowania}

%%%%%%%%%%%%%%%%%%%%%%%%%%%%%%%%%%%%%%%%%%%%%%%%%%%%%%%%%%%%%%%%%%%%%%%%%%%%%%%%
Pragn"e wyrazi"c g"l"ebok"a wdzi"eczno"s"c profesorowi Janowi
Kwieci"nskiemu,\\ kt"ory wprowadzi"l mnie w "swiat fizyki spinu, 
inspirowa"l i wspiera"l moj"a aktywno"s"c na tym polu. Bardzo dziekuj"e
profesor Barbarze Bade"lek za interesuj"ac"a i owocn"a wsp"o"lprac"e.
Kolegom z Zak"ladu Fizyki Teoretycznej Instytutu Fizyki J"adrowej jestem
wdzi"eczna za stworzenie mi"lej i stymuluj"acej atmosfery do pracy.
Bardzo dzi"ekuj"e\\ dr Piotrowi Czerskiemu za "ryczliw"a pomoc i cz"este
konsultacje komputerowe.

Dzi"ekuj"e r"ownie"r grupom: biologii strukturalnej profesora Janosa Hajdu
i fizyki wysokich energii profesora Gunnara Ingelmana na Uniwersytecie
w Uppsali za stworzenie doskona"lych warunk"ow do pracy, a tak"re za swobodny
dost"ep do szybkich jednostek obliczeniowych.

Bardzo dziekuj"e mojej rodzinie.

Badania przedstawione w tej rozprawie by"ly cz"e"sciowo finansowane przez
granty KBN Nr 2P03B 184 10, 2P03B 89 13, 2P03B 04214, 2P03B 05119,
2P03B 14420, 2P03B 04718, grant Wsp"olnoty Europejskiej "Training and Mobility
of Researchers" w ramach sieci "Quantum Chromodynamics and the Deep Structure
of Elementary Particles", FMRX-CT98-0194 oraz przez fundacje S. Batorego,
Wenner-Gren i stypendium habilitacyjne Instytutu Fizyki J"adrowej w Krakowie.

%%%%%%%%%%%%%%%%%%%%%%%%%%%%%%%%%%%%%%%%%%%%%%%%%%%%%%%%%%%%%%%%%%%%%%%%%%%%%%%%

\section*{Acknowledgements}

%%%%%%%%%%%%%%%%%%%%%%%%%%%%%%%%%%%%%%%%%%%%%%%%%%%%%%%%%%%%%%%%%%%%%%%%%%%%%%%%
I would like to express my deep gratitude to Professor Jan Kwieci"nski for
introducing me into the fascinating world of the spin physics,
for inspiring and supporting my activity in this field. I thank Professor 
Barbara Bade"lek for the
interesting and fruitful collaboration. I am grateful to my colleagues from
the Department of Theoretical Physics at the Institute  of Nuclear Physics
in Krak"ow for creating a good atmosphere to work. I thank Dr. Piotr Czerski
for the computer support.

I am grateful to the groups of Professor Janos Hajdu (Department of
Biochemistry) and Professor Gunnar Ingelman (ISV, High Energy Physics)
at the Uppsala University for providing me with the excellent conditions
to work. I am especially grateful for having a free access to fast
computational units.

I thank my family.

This research was supported in part by the grant of the Polish KBN grants
Nos. 2P03B 184 10, 2P03B 89 13, 2P03B 04214, 2P03B 05119,
2P03B 14420, 2P03B 04718, and the grant of the European Community ''Training and
Mobility of Researchers'' at the Network "Quantum Chromodynamics and the Deep
Structure of Elementary Particles", FMRX-CT98-0194. I was also supported by
the S. Batory Foundation, the Wenner-Gren Foundation and the habilitation
scholarship of the Institute of Nuclear Physics in Krak"ow.

%%%%%%%%%%%%%%%%%%%%%%%%%%%%%%%%%%%%%%%%%%%%%%%%%%%%%%%%%%%%%%%%%%%%%%%%%%

\section*{Dodatek A}

%%%%%%%%%%%%%%%%%%%%%%%%%%%%%%%%%%%%%%%%%%%%%%%%%%%%%%%%%%%%%%%%%%%%%%%%%%
Podajemy tu kr"otk"a charakterystyk"e  j"ader ewolucji w r"ownaniu \re{nloinf}.
J"adra ewolucji DGLAP podajemy za Ref.\ \cite{reya}. Pe"lne j"adro ewolucji
DGLAP: $\dt P$ zawiera zar"owno cz"lony wiod"ace, jak i podwiod"ace:
\eq
\dt P=\dt P_{LO} + \as\, \dt P_{NLO}.
\label{dglap}
\eqx
Do cz"lonu jednorodnego $\displaystyle \as\int_{k_0^2}^{Q^2} {dk^2 \over k^2}
\,(\dt P_{reg}\ot f_{NS})(x,k^2)$ w r"ownaniu \re{nloinf}, w"l"aczamy tylko
cz"e"s"c regularn"a pe"lnego j"adra ewolucji. Czynimy tak, aby unikn"a"c
podw"ojnego liczenia wk"lad"ow logarytmicznych $\ln^2(1/x)$
pochodz"acych z cz"lon"ow
NLO DGLAP i z cz"lon"ow pozadrabinkowych w tym samym
obszarze przestrzeni fazowej: $k_0^2<k^2<Q^2$.

\noindent
J"adra ewolucji $\ln^2(1/x)$ w obszarze $Q^2<k^2<Q^2/z$,
uzyskane z resumacji diagram"ow drabinkowych, odpowiadaj"a
j"adrom ewolucji DGLAP
przy zerowym u"lamku p"edu pod"lu"rnego $z=0$ \pc{I}.

\noindent
J"adra ewolucji pozadrabinkowej zosta"ly otrzymane w pracy \pc{I}
z podczerwonych
r"owna"n ewolucji dla singletowych fal cz"astkowych ${\bf F_0}$, $\bf F_8$
\cite{BARTNS,BARTS,QCD,QCD1}. W \pc{I} zauwa"ryli"smy, "re rozszerzenie j"adra
ewolucji $\ln^2(1/x)$:
\eq
\as \dt P_{qq}/\omega\nonumber,
\eqx
do zmodyfikowanej postaci:
\eq
\as \left( \dt P_{qq}/\omega
-({\bf F_8}(\omega)\,{\bf G_0})_{qq}/(2\pi^2\omega^2) \right)\nonumber,
\eqx
prowadzi"lo do odtworzenia wymiaru anomalnego,
wygenerowanego poprzez podczerwone r"ownania ewolucji.

\noindent
Macierz ${\bf G}_0$ zawiera czynniki kolorowe reprezentuj"ace przy"l"aczenie
mi"ekkich gluon"ow do zewn"etrznych n"og amplitudy rozpraszania:
\eqn
{\bf G}_0 &=&\left( \begin{array}{cc}  {N^2-1 \over 2N} & 0  \\
                                                    0 & N  \\ \end{array}
						    \right ),
\label{g0}
\eqnx
gdzie $N$ oznacza liczb"e kolor"ow.

\noindent
Sprawdzili"smy, "re zastosowanie przybli"renia Borna do fali cz"astkowej,
${\bf F_8}$,
\eq
{\bf F}_8^{Born}(\omega)\approx 8\pi^2 \as \frac{{\bf M}_8}{\omega}.
\eqx
dawa"lo wystarczaj"aco dok"ladne oszacowanie pozadrabinkowej
ewolucji $\ln(1/x)$.
Macierz ${\bf M}_8$ oznacza"la macierz j"ader ewolucji w oktetowym kanale
$t$:
\eqn
{\bf M}_8 &=&\left( \begin{array}{cc} -{1 \over 2N} & -{N_F \over 2}\\
                                                N & 2N \\ \end{array} \right ).
\eqnx
Odwrotna transformata Mellina wielko"sci ${\bf F}_8^{Born}(\omega)$
mia"la wtedy posta"c:
\eq
\Biggl[\frac{\tilde {\bf F}_8^{Born}}{\omega^2}\Biggr](z)=
4\pi^2 \as {\bf M}_8 ln^2 (z).
\label{born}
\eqx

\noindent
R"ownanie ewolucji \re{nloinf} opisuje cz"e"s"c pozadrabinkow"a ewolucji
$\ln^2(1/x)$ w przybli"reniu Borna  \re{born}.
%%%%%%%%%%%%%%%%%%%%%%%%%%%%%%%%%%%%%%%%%%%%%%%%%%%%%%%%%%%%%%%%%%%%%%%%%%

%%%%%%%%%%%%%%%%%%%%%%%%%%%%%%%%%%%%%%%%%%%%%%%%%%%%%%%%%%%%%%%%%%%%%%%%%%%%%%%%

% LITERATURA

%%%%%%%%%%%%%%%%%%%%%%%%%%%%%%%%%%%%%%%%%%%%%%%%%%%%%%%%%%%%%%%%%%%%%%%%%%%%%%%%
%\bibliographystyle{unsrt}
%\bibliography{mom1}

%%%%%%%%%%%%%%%%%%%%%%%%%%%%%%%%%%%%%%%%%%
\end{document}